\documentclass[aps,prb,reprint,showpacs,superscriptaddress]{revtex4-1}
\usepackage{color}
\usepackage{pict2e}
\usepackage{graphicx}

\newcommand{\eqn}[1]    {(\ref{#1})}
\newcommand{\fig}[1]	{Figure \ref{#1}}

\def\etal	{{\em et al.\ }}

\def\bi         {\begin{itemize}}
\def\ei         {\end{itemize}}
\def\benu	{\begin{enumerate}}
\def\eenu	{\end{enumerate}}
\def\bmat       {\left[ \begin{array}}
\def\emat       {\end{array} \right]}
\def\beq	{\begin{equation}}
\def\eeq	{\end{equation}}
\def\beqn       {\begin{eqnarray*}}
\def\eeqn       {\end{eqnarray*}}
\def\beqa       {\begin{eqnarray}}
\def\eeqa       {\end{eqnarray}}
\def\bquote	{\begin{quote}}
\def\equote	{\end{quote}}

\def\bwide	{\begin{widetext}}
\def\ewide	{\end{widetext}}

\def\s          {\sigma}

\def\dag	{\dagger}

\def\im		{{\mbox{Im}}}
\def\re		{{\mbox{Re}}}

\begin{document}


\title{Analytic approach to the edge state of the Kane-Mele Model}

\author{Hyeonjin Doh}
\affiliation{Department of Physics and
Center for Computational Studies of Advanced Electron Material Properties,
Yonsei University, Seoul 120-749, Korea}
\thanks{This work was supported by NRF of Korea (Grant No. 2011-0018306).}

\author{Gun Sang Jeon}
\affiliation{Department of Physics, Ewha Womans University, Seoul 120-750, Korea}
\thanks{This work was supported by NRF of Korea (Grant No. 2008-0061893).}

\author{Hyoung Joon Choi}
\email[]{h.j.choi@yonsei.ac.kr}
\affiliation{Department of Physics and
Center for Computational Studies of Advanced Electron Material Properties,
Yonsei University, Seoul 120-749, Korea}


\date{\today}

\begin{abstract}
We investigate the edge state of a two-dimensional topological insulator 
based on the Kane-Mele model.
Using complex wave numbers of the Bloch wave function,
we derive an analytical expression for the edge state localized near the edge of 
a semi-infinite honeycomb lattice with a straight edge.
For the comparison of the edge type effects,
two types of the edges are considered in this calculation;
one is a zigzag edge and the other is an armchair edge.
The complex wave numbers and the boundary condition give
the analytic equations for the energies and the wave functions
of the edge states.
The numerical solutions of the equations reveal
the intriguing spatial behaviors of the edge state.
We define an edge-state width for analyzing the spatial variation of the edge-state
wave function. 
Our results show that the edge-state width can be easily controlled by a couple of
parameters such as the spin-orbit coupling and the sublattice potential.
The parameter dependences of the edge-state width show substantial differences
depending on the edge types.
These demonstrate that, even if
the edge states are protected by the topological property of the bulk,
their detailed properties are still discriminated by their edges.
This edge dependence can be crucial in manufacturing small-sized
devices since the length
scale of the edge state is highly subject to the edges.
\end{abstract}


\maketitle

\section{Introduction}
Since the Hall coefficient of the integer quantum Hall effect (IQHE) 
was known to be described with the topological index,\cite{Thouless1982prl}
the topological quantum state has become one of the main branches in 
condensed matter physics.
Due to the different topological nature of this state, there should exist a
gapless mode at the boundary or at the interface even if the bulk state is 
energetically gapped.
There have been lots of efforts to find the similar topological states without 
a magnetic field.
Murakami, Nagaosa, and Zhang found that spin-orbit coupling (SOC) can play the role of
the magnetic field in IQHE, and dubbed it as a quantum spin Hall effect
(QSHE).\cite{Murakami2003} 
Kane and Mele suggested a specific model Hamiltonian which shows
QSHE.\cite{Kane2005,Kane2005a} 
Soon after that, Bernevig et al.\cite{Bernevig2006,Bernevig2006a} showed that
QSHE can be realized in
two-dimensional HgTe/CdTe quantum well,
and it was confirmed by experiments.\cite{Konig2007,Konig2008}
The system showing QSHE is also dubbed as a topological insulator (TI) after
these works were extended to three-dimensional (3D) materials.\cite{Moore2007,Fu2007}
The 3D TIs are also confirmed by
angle resolved photoemission spectroscopy (ARPES) experiments
in Bi$_x$Sb$_{1-x}$,\cite{Hsieh2008} Bi$_2$Se$_3$,\cite{Xia2009} and
Bi$_2$Te$_3$.\cite{Chen2009,Hsieh2009}
Consistent results are also yielded
by the scanning tunneling microscope (STM) measurements
with Bi$_{1-x}$Sb$_x$\cite{Roushan2009} and Bi$_2$Te$_3$.\cite{Alpichshev2010}

A series of the successes in the TI now open a new route to
the novel phases such as Weyl semi-metal and  Majorana fermion state.\cite{Wan2011} 
The interests in this area are also extended to
the application-related science such as
quantum computing science and spintronics  due to 
the topologically protected surface state 
and its spin chirality from the time-reversal symmetry.

Despite the success of the ARPES and the STM experiments,
the transport experiments of the TI are not
so successful to show the metallic surface state.
Current TIs are mostly naturally doped or have
imperfections in the bulk.
This generates residual bulk carriers which
prevent the surface current from being detected separately.
One of the efforts to reduce the residual carriers is to decrease the thickness of the
sample. 
However, the thickness-dependent experiments show a gap opening when the 
film is thin enough to cause the overlap between the wave functions on
the two opposite surfaces.\cite{Zhang2010}
To prevent the overlap of the edge states,
we need to control the spatial variation of the edge state in the bulk.
Especially, this spatial dependence of the edge-state wave function 
can be crucial when we consider a small system as in the device manufacturing with
TIs.

In this paper, we develop an analytic approach
to the Kane-Mele model for the microscopic understanding of the surface state of
the TI.
Especially, we focus on the spatial profiles of the edge state in the bulk region.
First, in Sec.\ \ref{sec:selfeqn}, we extend the analytic approach for
the edge-state wave function developed
by K\"{o}nig \etal\cite{Konig2008} and Wang \etal\cite{Wang2009}
to the Kane-Mele model.
The effect of the edge type is also investigated by
considering two typical types of the edge in the honeycomb lattice;
an armchair (AC) edge and a zigzag (ZZ) edge.
Therefore, we derive two sets of self-consistent equations for the two edges,
respectively.
In Sec.\ \ref{sec:dispersion},
we solve the self-consistent equations numerically,
but we also derive an analytical expression for the special case such as
$k=\pi$ for the ZZ edge
or the case without the sublattice potential, $\lambda_v=0$.
These results show significant distictions
between the systems with the different edges.
In this section, we focus on the effect of the internal and the external parameters
on the energy dispersion of the edge state, and how the effects are different
depending on the edge.
The profiles of the edge-state wave function will be discussed in Sec.\ \ref{sec:width}. The spatial properties of the edge state will be
discussed by defining the width of the edge state which shows the length scale
of the  wave function decaying into the bulk.
The parameter-dependent edge-state width will be discussed further
for the controlling of the edge-state gap in a finite sized system
in Sec.\ \ref{sec:finite}.
Finally, we will summarize the results in Sec.\ \ref{sec:summary}.

\section{Self-consistent equation for Kane-Mele model in semi-infinite lattice}
\label{sec:selfeqn}
In this section we will derive the self-consistent equations for the edge state of 
Kane-Mele (KM) model in the semi-infinite lattice.
First, we construct the Harper's equation\cite{Harper1955}
which describes the wave function in real space
in the direction normal to the edge.
We consider complex momenta of the Bloch wave function in the direction normal
to the edge, since the translational symmetry is broken in that direction.
The complex momentum gives the decaying wave function of the edge state. 
Next, the Harper's equation
yields the effective Hamiltonian for the edge state by using the decaying
wave function.
Finally, we consider boundary conditions of the edge state, deriving a complete set of 
equations for the edge-state energy dispersion and the decaying factors of the edge
state.

We start from the Kane-Mele model\cite{Kane2005, Kane2005a}
which is the tight-binding model
with a spin-orbit interaction in a honeycomb lattice.
\begin{eqnarray}
	H&=&-t\sum_{\langle i,j\rangle\sigma}
	c_{i\sigma}^\dagger c_{j\sigma}^{}
	+i\lambda_{SO}\!\!\!\!\sum_{\langle\langle i,j\rangle\rangle\alpha,\beta}
	\nu_{ij}\sigma_{\alpha\beta}^{z}c_{i\alpha}^\dagger c_{j\beta}^{}
	\nonumber \\
	&&+\lambda_v\sum_{i\s} \zeta_i c_{i\sigma}^\dagger c_{i\sigma}^{}.
	\label{eq:KMmodel}
\end{eqnarray}
Here, the lattice index $i$ and $j$ are the site indices of the honeycomb lattice.
$\lambda_{SO}$ is a spin-orbit coupling strength through
the next-nearest-neighbor hopping
and $\nu_{ij}=\pm 1$ is determined by 
$\nu_{ij} = \frac{2}{\sqrt{3}}(\mathbf{d}_1\times\mathbf{d}_2)$ where
$\mathbf{d}_{1,2}$ are the adjecent two vectors denoting the two nearest-neighbor bonds
connecting the next-nearest-neighbor sites.
$\sigma^z$ is the $z$ component of Pauli matrix, and
$\lambda_v$ is the sub-lattice potential with $\zeta_i=\pm1$.
In the momentum space, the Hamiltonian can be written as
\beqa
H = 4\lambda_{SO}\sin q_x (\cos q_x -\cos q_y)\Gamma^{15} -\lambda_v\Gamma^2 
\nonumber \\
-t\left(2\cos q_x + \cos q_y\right)\Gamma^1
-t\sin q_y \Gamma^{12}
\label{eq:hamiltonian_q}
\eeqa
where $q_x=\frac{k_x}{2}$, and $q_y=\frac{\sqrt{3}}{2}k_y$.
Here, $\Gamma^{ab}$ and $\Gamma^a$ are Dirac matrices shown
in Table \ref{table:Dirac}.
The energy spectrum can be acquired by diagonalizing
the Hamiltonian \eqn{eq:hamiltonian_q}.
\beqa
E_{{\bf q}\sigma}^2 = \left\{4\lambda_{SO}\sin q_x\left(\cos q_x-\cos q_y\right)
-\sigma\lambda_v\right\}^2
\nonumber \\
+t^2\left\{
1+4\cos^2q_x+4\cos q_x\cos q_y
\right\}
\eeqa
which gives the bulk energy gap $|3\sqrt{3}\lambda_{SO}\pm\lambda_v|$
at $K$ and $K'$ in the Brillouin zone of the honeycomb lattice.
The $\pm$ sign in the gap depends on the position $K$, $K'$  and electron spin.
Although both $\lambda_{SO}$ and $\lambda_v$ can generate a bulk gap separately,
they give topologically different characters to the gap;
The $\lambda_{SO}$ gives a topologically non-trivial gap
and the $\lambda_v$ does a trivial gap.
Therefore, when we consider the two parameters at the same time,
they compete with each other, 
and the system can be gapless with a proper ratio of the parameters.
Actually, the system has a transition between
the topologically trivial state to
the non-trivial state when $\lambda_v=3\sqrt{3}\lambda_{SO}$.
If the system becomes the topologically non-trivial state, 
it has a gapless edge state on its edge.
To investigate the edge state, we need to set an edge in the system.
In this work, we will consider
two typical edges of a honeycomb lattice; ZZ and AC edges.

\begin{table}[tb]
\begin{tabular}{||c|c|c|c|c||}
	\hline
$\Gamma^1$ & $\Gamma^2$ & $\Gamma^3$ & $\Gamma^4$ & $\Gamma^5$ \\
$\tau^x\otimes I$ & $\tau^z\otimes I$ & $\tau^y\otimes \sigma^x$ &
$\tau^y\otimes \sigma^y$ & $\tau^y\otimes \sigma^z$  \\ \hline
\multicolumn{5}{||c||}{$ 
\Gamma^{ab} = \left[\Gamma^a,\Gamma^b\right]/(2i)
$} \\ \hline
\end{tabular}
\caption{
\label{table:Dirac}
Dirac matrices where $\tau$ and $\sigma$ are the Pauli matrices represented
in the sublattice space and the spin space, respectively.
The operator $\otimes$ is the Kronecker matrix product.}
\end{table}

\begin{figure}[tbh]
	\includegraphics[width=8cm]{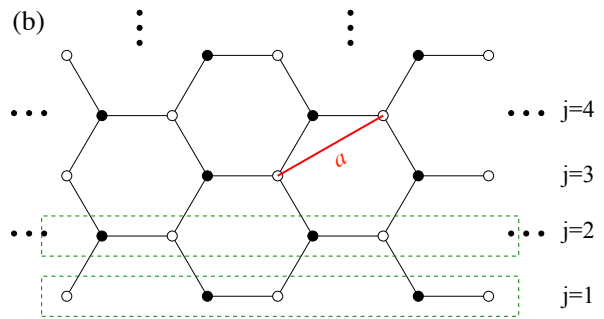}
	\caption{
	\label{fig:latticeAC}
	The semi-infinite honeycomb lattice of a lattice constant $a$ 
	with an AC edge. Here, $j$ is a real-space row index in $y$-direction 
	perpendicular to the edge.}
\end{figure}
\subsection{Semi-infinite lattice with an armchair  edge}
\subsubsection{Harper's equation}

As shown in \fig{fig:latticeAC},
we first consider a semi-infinite lattice
which covers the upper-half region ($y>0$) in two-dimensional space and
has an AC edge along $x$-axis.
In this case,
the Hamiltonian \eqn{eq:KMmodel} can be written in the momentum $k$ 
in $x$-direction 
and in the real space lattice index $j$ in $y$-direction.
Then the Hamiltonian can be written as
\beqa
H&=&\sum_{qj}\left(
\Psi_{qj}^\dag\hat{M}_{AC}\Psi_{qj}^{}
+\Psi_{qj+1}^\dag\hat{T}_{AC}\Psi_{qj}
+\Psi_{qj}^\dag\hat{T}_{AC}^\dag\Psi_{qj+1}\right. \nonumber \\
&&\left.+\Psi_{qj+2}^\dag\hat{T}'_{AC}\Psi_{qj}
+\Psi_{qj}^\dag\hat{T}'^\dag_{AC}\Psi_{qj+2}\right)
\label{eqn:hamiltonian_ac}
\eeqa
where
\beqa
\hat{M}_{AC}&=&
-\lambda_v \Gamma^2
-t \Gamma^1,
\\
\hat{T}_{AC}&=&
2i\lambda_{SO}\cos q ~ \Gamma^{15}
-t\cos q  ~ \Gamma^1
-t\sin q  ~ \Gamma^{12},
\\
\hat{T}'_{AC}&=&
-i\lambda_{SO} \Gamma^{15},
\eeqa
and $q$ is defined as 
$q = \frac{\sqrt{3}}{2}k$.
Now, we can construct 
the Harper's equation for the two-dimensional wave function $\Psi_{qj}$
in the following forms, 
\beqa
E_q\Psi_{qj} &=&\hat{M}_{AC}\Psi_{qj}
+\hat{T}_{AC}\Psi_{qj-1}
+\hat{T}_{AC}^\dag\Psi_{qj+1}
\nonumber \\
&&+\hat{T}'_{AC}\Psi_{qj-2}
+\hat{T}_{AC}'^\dag\Psi_{qj+2},
\label{eqn:ac_harpers}
\eeqa
where $E_q$ is the energy eigenvalue of the Hamiltonian \eqn{eqn:hamiltonian_ac}
for the momentum $q$.

\subsubsection{Effective Hamiltonian for the edge state}

Now, we allow a complex momentum $i\kappa$
for the Bloch wave function in $y$-direction,\cite{Konig2008,Wang2009} 
which can provide a solution to the system with a boundary.
The complex momentum eventually makes the wave function
 decay  as it gets into the bulk.
Therefore, we can simplify the $j$ dependence of the wave function
by using the complex momentum $\kappa$ 
as the following form 
\beq
\Psi_{qj} = e^{-\kappa (j-1)}\Psi_{q}.
\label{eqn:decaying_factor}
\eeq
Here, the complex momentum $\kappa$ means the decaying factor
of the wave function and it is generally a complex number whose real part is positive
for the current boundary condition.
The effective Hamiltonian for the edge state can be written with the decaying
wave function as
\beq
E_q\Psi_q=\hat{H}_{AC}\Psi_q,
\eeq
where
\beq
\hat{H}_{AC}=\hat{M}_{AC}
+ e^{\kappa}\hat{T}_{AC} + e^{-\kappa}\hat{T}_{AC}^\dag
+ e^{2\kappa}\hat{T}'_{AC} + e^{-2\kappa}\hat{T}_{AC}'^\dag.
\eeq
The explicit matrix form of the effective Hamiltonian for spin $\sigma$ is
\begin{widetext}
\beq
\hat{H}_{AC}^{\mbox{\scriptsize edge}}=\bmat{cc}
4i\sigma\lambda_{SO}\sinh\kappa\left(\cosh\kappa-\cos q\right)-\sigma\lambda_v &
-t\left(1+2e^{-iq}\cosh\kappa\right) \\
-t\left(1+2e^{iq}\cosh\kappa\right) &
-4i\sigma\lambda_{SO}\sinh\kappa\left(\cosh\kappa-\cos q\right)+\sigma\lambda_v 
\emat.
\label{eqn:ac_hamiltonian}
\eeq
From this Hamiltonian, we can get the following eigenvalue equation,
\beq
E_q^2=
\left\{4i\lambda_{SO}\sinh\kappa\left(
\cosh\kappa - \cos q\right)
-\sigma\lambda_v\right\}^2
+t^2\left\{1+4\cos q\cosh\kappa+4\cosh^2\kappa\right\}.
\label{eqn:ac_energy}
\eeq
\end{widetext}
For a given momentum $q$ and energy  $E_q$, this eigenvalue equation always yields
four solutions for the decaying factor, $\kappa$.
Thus, with this equation only, we cannot determine 
the energy dispersion of the edge state, that is, the $q$-dependence of $E_q$.
For the complete set of equations,
we should also consider boundary conditions for the edge-state energy dispersion, 
which we will derive in the following section.

\subsubsection{Boundary condition of the edge state}
Let $\kappa_\nu$ is the $\nu$-th solution of the eigenvalue equation \eqn{eqn:ac_energy}
for a given $q$ and a given energy $E_q$.
Then the eigenvector for the value of $q$, the energy $E_q$,
and its solution $\kappa_\nu$
can be written as
\beq
\Phi_{\nu\sigma}^{AC}(q)
=\bmat{c}
t\left(1+2e^{iq}\cosh\kappa_\nu\right) \\
4i\sigma\lambda\sinh\kappa_\nu\!\left(\cosh\kappa_\nu\!-\!\cos q\right)
\!-\!\lambda_v\!-\!E_q
\emat.
\label{eqn:ac_vector}
\eeq
Here, $\nu(=1,\cdots,4)$ is an index of the solutions.
The general solution, satisfying $\lim_{j\rightarrow\infty}\Psi_{q}=0$,
can be written as a linear combination of the eigenvectors.
\beq
\Psi_{q\sigma}(j)
=\sum_\nu a_{q\nu} e^{-\kappa_\nu(j-1)}
\Phi_{\nu\sigma}^{AC}(q),
\label{eqn:general_sol}
\eeq
where $a_{q\nu}$s are arbitrary constants and should be determined
by the boundary condition of the system.
Now the boundary condition requires $\Psi_k(j\leq0)=0$.
From the Harper equation \eqn{eqn:ac_harpers},
it can be shown that
the boundary condition 
requires the following condition,
\beq
\Psi_{q}(j=0) = \Psi_{q}(j=-1)=0,
\label{eqn:ac_bc0}
\eeq
since the Haper's equation \eqn{eqn:ac_harpers}
contains the coupling between $\Psi_{k}(j)$ and $\Psi_{k}(j+2)$.
\beq
\bmat{cccc}
\Phi_{q1\sigma}^{AC} & \Phi_{q2\sigma}^{AC} &
\Phi_{q3\sigma}^{AC} & \Phi_{q4\sigma}^{AC} \\
e^{-\kappa_1}\Phi_{q1\sigma}^{AC} & e^{-\kappa_2}\Phi_{q2\sigma}^{AC} & 
e^{-\kappa_3}\Phi_{q3\sigma}^{AC} & e^{-\kappa_4}\Phi_{q4\sigma}^{AC}
\emat
\bmat{c}
a_{q1} \\ a_{q2} \\ a_{q3} \\ a_{q4}
\emat
=0
\eeq
To have a non-trivial solution for $a_{q\nu}$
we should have
\beq
\left|\begin{array}{cccc}
	\Phi_{q1\sigma}^{AC} & \Phi_{q2\sigma}^{AC} &
	\Phi_{q3\sigma}^{AC} & \Phi_{q4\sigma}^{AC} \\
	e^{-\kappa_1}\Phi_{q1\sigma}^{AC} & e^{-\kappa_2}\Phi_{q2\sigma}^{AC} & 
	e^{-\kappa_3}\Phi_{q3\sigma}^{AC} & e^{-\kappa_4}\Phi_{q4\sigma}^{AC}
\end{array}\right|=0.
\label{eqn:ac_bc}
\eeq

The eigenvalue equation
\eqn{eqn:ac_energy} and
the boundary condition
\eqn{eqn:ac_bc}
provide
a complete set of equations
for the edge-state energy dispersion and the decaying factors
of the edge state.
The solutions from the coupled equations can be obtained numerically
by an iterative method described in 
Sec.\ \ref{sec:dispersion} and
the obtained solutions are presented and analyzed in
Secs.\ \ref{sec:dispersion} and \ref{sec:width}.

\begin{figure}[tbh]
	\includegraphics[width=8cm]{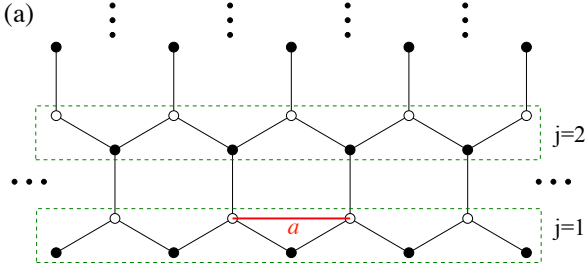} \\
	\caption{
	\label{fig:latticeZZ}
	The semi-infinite honeycomb lattice of a lattice constant $a$
	with a ZZ edge.
	Here, $j$ is a real-space row index in $y$ direction perpendicular to the edge.}
\end{figure}
\subsection{Semi-infinite lattice with a zigzag  boundary}

For the zigzag boundary, we already derived the equations for the energy
and the wave function in our previous work\cite{Doh2013prb}.
In this section, we will just summarize the
derivation with the new notations in this paper. 
Based on the semi-infinite lattice as shown in \fig{fig:latticeZZ}
the Hamiltonian with the momentum $k$ in $x$-direction
and real space index $j$ in $y$-direction can be written
as
\beq
H=\sum_{qj}\left(
\Psi_{qj}^\dag\hat{M}_{ZZ}\Psi_{qj}^{}
+\Psi_{qj+1}^\dag\hat{T}_{ZZ}\Psi_{qj}
+\Psi_{qj}^\dag\hat{T}_{ZZ}^\dag\Psi_{qj+1}\right)
\eeq
where
\beqa
\hat{M}_{ZZ}&=&2\lambda_{SO}\sin 2q ~ \Gamma^{15}
-2t\cos q ~ \Gamma^1
-\lambda_v\Gamma^2,
\\
\hat{T}_{ZZ} &=& 
-2\lambda_{SO}\sin q  ~ \Gamma^{15}
-t(\Gamma^1-i\Gamma^{12}).
\eeqa
Here, $q$ is defined as $q=\frac{k}{2}$ in the parallel direction to the edge.
Like the previous section, the Harper's equation is
\beq
E_q\Psi_{qj} = \hat{M}_{ZZ}\Psi_{qj}
+\hat{T}_{ZZ}\Psi_{qj-1}
+\hat{T}_{ZZ}^\dag\Psi_{qj+1}.
\label{eqn:zz_harpers}
\eeq
With the same decaying form of the wave function in \eqn{eqn:decaying_factor},
the effective Hamiltonian can be written as
\beq
\hat{H}_{ZZ}=\hat{M}_{ZZ} + e^{\kappa} \hat{T}_{ZZ} + e^{-\kappa} \hat{T}_{ZZ}^\dag.
\eeq
The explicit form of the Hamiltonian for the electron with spin $\sigma$ is
\begin{widetext}
\beq
\hat{H}_{ZZ}=\bmat{cc}
	4\sigma\lambda_{SO}\sin q\left(\cos q-\cosh\kappa\right)-\sigma\lambda_v &
	-t\left(2\cos q+\cosh\kappa-\sinh\kappa\right) \\
	-t\left(2\cos q+\cosh\kappa+\sinh\kappa\right) &
	-4\sigma\lambda_{SO}\sin q\left(\cos q-\cosh\kappa\right)+\sigma\lambda_v 
	\emat.
\label{eqn:zz_hamiltonian} 
\eeq
From this Hamiltonian, we can get the following eigenvalue equation,
\beq
E_q^2=
\left\{4\lambda_{SO}\sin q\left(\cos q-\cosh\kappa\right)
-\sigma\lambda_v\right\}^2
+t^2\left\{4\cos^2 q+ 4\cos q\cosh\kappa + 1\right\},
\label{eqn:zz_energy}
\eeq
\end{widetext}
which gives two values of $\kappa$ for  given $q$ and $E_q$.
Now, we can write down the following $\nu$-th eigenvector
corresponding to
the solution $\kappa_\nu$,
\beq
\Phi_{\nu\sigma}^{ZZ}
=\bmat{c}
t\left(2\cos q\!+\cosh\kappa_\nu+\sinh\kappa_\nu\right) \\
4\sigma\lambda_{SO}\sin q\!\left(\cos q\!-\!\cosh\kappa_\nu\right)
\!-\!\lambda_v\!-\!E_q
\emat.
\label{eqn:zz_vector}
\eeq
From the Harper's equation of the ZZ edge \eqn{eqn:zz_harpers},
it can be shown that
the boundary condition can be satisfied
by 
\beq
\Psi_{k}(j=0) = 0.
\eeq
This means that the two eigenvectors,
$\Phi_{1\sigma}^{ZZ}$ and $\Phi_{2\sigma}^{ZZ}$
should be linearly dependent.
Therefore, if we define a $2\times 2$ matrix composed of the two vectors, its 
determinant should be zero 
\beq
\left|\begin{array}{cc}
	\Phi_{q1\sigma} & \Phi_{q2\sigma}
\end{array}
\right|=0.
\label{eqn:zz_bc}
\eeq
Again we express this boundary condition
in terms of the eigenvectors \eqn{eqn:zz_vector}
which are also functions of $q$, $E_q$, and the decaying factors $\kappa$s.
The solution of the two coupled equations for ZZ edge, \eqn{eqn:zz_energy}
and \eqn{eqn:zz_bc}, will
be presented in the following two sections along with those of the AC edge.

\begin{figure}[t]
	\includegraphics[width=8cm]{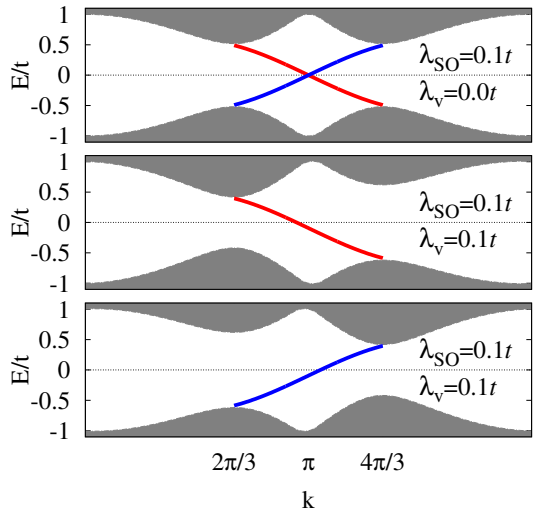}
	\caption{\label{fig:zz_dispersion}
	The edge-state energy dispersion of the ZZ edge for different
	parameters of the SOC and the sub-lattice potential.
	The gray shadowed regions are the bulk energy spectra. 
	The red  line is for the spin-up electron,
	and the blue line for the spin-down electron.
	The topmost graph is the dispersion of the edge state and the energy spectra of
	the bulk state without any sublattice potential and the lower two 
	are those with a sublattice potential of $\lambda_v=0.1t$
	for spin-up and spin-down electrons, respectively.
	}
\end{figure}

\section{Energy spectrum}
\label{sec:dispersion}
In this section, we solve the coupled equations,
\eqn{eqn:zz_energy} and \eqn{eqn:zz_bc} for the edge-state
energy of the ZZ edge,
and \eqn{eqn:ac_energy} and \eqn{eqn:ac_bc} for that of the AC edge.
The overall results in this section
generally agree with
the previous numerical
results.\cite{Kane2005,Ezawa2013prb}
Nevertheless, since we deal with a semi-infinite lattice,
our results are free of any numerical errors which may come from the 
small size of the system.

For a given $q$, we solve the coupled equations iteratively
by the following steps.
First, from the initial value of the energy, we solve the equation
of $\kappa$s, \eqn{eqn:zz_energy} for the ZZ edge
and \eqn{eqn:ac_energy} for the AC edge.
From the solutions, we write down the eigenvectors \eqn{eqn:zz_vector}
and \eqn{eqn:ac_vector}.
Finally, as shown in \eqn{eqn:zz_bc} and \eqn{eqn:ac_bc},
we construct the $2\times 2$ matrix ($4\times 4$ matrix)
with the two (four) eigenvectors from
\eqn{eqn:zz_vector} for the ZZ boundary (from \eqn{eqn:ac_vector} for the AC boundary),
and extract the new energy value from its determinant condition.
Now the new energy is plugged in the first step and these steps are repeated until
the energy is converged.

\fig{fig:zz_dispersion} shows the energy dispersion of the edge state for the
ZZ boundary.
The bulk energy gap occurs at $k=\frac{2\pi}{3}$ and $\frac{4\pi}{3}$ which
correspond to the $K$ and $K'$ points of the two-dimensional
Brillouin zone of the honeycomb lattice.
The dispersion of the edge state crosses the gap
and connects the valence band maximum at $K$ and the conduction band
minimum at $K'$, and vice versa for the opposite spin.
Without the sub-lattice potential,
the spin-up and spin down dispersions cross at $k=\pi$ and
the center of the bulk energy gap.
If we expand the self consistent equations \eqn{eqn:zz_energy}, \eqn{eqn:zz_vector},
and \eqn{eqn:zz_bc} for small $k-\pi$ and $\lambda_v=0$,
then we get the following expression for 
the energy near $k=\pi$.
\beq
E_{k}^{ZZ} \simeq v_F^{ZZ} (k-\pi)
\eeq
where $v_F^{ZZ}$ is the Fermi velocity of the half-filled system with $\lambda_v=0$.
\beq
\label{eqn:FermiVelocity}
v_F^{ZZ} = \pm\frac{6\lambda_{SO}t}{\sqrt{t^2+16\lambda_{SO}^2}}
\eeq
Here, the Fermi velocity is linearly proportional to the SOC for small SOC,
and its derivation is shown in Appendix \ref{sec:zz_Fermi_vel}.

For the finite sub-lattice potential, the bulk energy spectrum is asymmetric around
$k=\pi$. Especially, one of the bulk gaps at $K$ and $K'$ get narrower, and the other
gets wider.
Therefore, the edge dispersion connecting two band edge at $K$ and $K'$,
moves vertically.
The energy shift at $k=\pi$ are proportional to $\lambda_v$ as derived
in our previous work\cite{Doh2013prb}
\beq
E_\pi^{ZZ} = \frac{\lambda_v t}{\sqrt{t^2+16\lambda_{SO}^2}}.
\eeq
One of the energy gaps is finally closed when $3\sqrt{3}\lambda_{SO}=\lambda_v$.
The further increase of the sub-lattice potential, $\lambda_v$, however,
will reopen the bulk energy gap accompanied by the edge-state energy-gap opening.

\begin{figure}[hbt]
	\includegraphics[width=8cm]{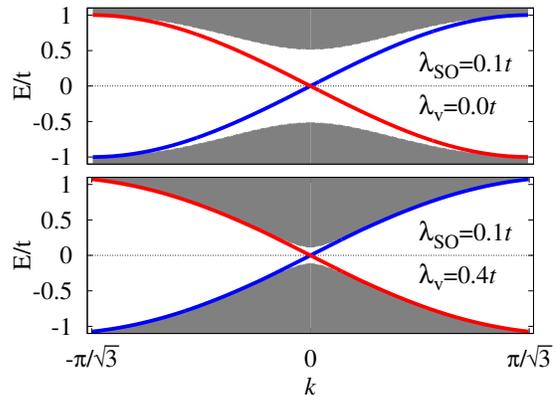}
	\caption{\label{fig:ac_dispersion}
	The edge-state
	energy dispersion of the AC edge for the SOC strength of 
	$\lambda_{SO}=0.1t$.
	The upper graph shows the energy dispersion without any sublattice potential,
	and the lower one, with the sublattice potential of $\lambda_v=0.4t$.
	The gray filled regions 
	are the bulk energy spectra.
	The red and blue solid lines are the edge-state dispersion relations for
	the spin-up and spin-down electrons, respectively.
	}
\end{figure}

\fig{fig:ac_dispersion} shows the edge-state energy dispersion
on the AC edge.
Without the sub-lattice potential, the edge-state energy dispersion can be
expressed as
\beq
E_k^{AC} = t \sin\frac{\sqrt{3}}{2}k,
\label{eqn:ac_dispersion}
\eeq
which is derived in Appendix \ref{sec:ac_energy_disp}.
Interestingly, the edge state dispersion is not affected by the SOC on the AC boundary.
The SOC modifies only the bulk energy spectrum by changing the bulk energy gap.
If we consider the sub-lattice potential, the gap is suppressed due to the 
competition with the SOC. 
Increasing the sub-lattice potential makes
the gap smaller until
the gap is closed at $\lambda_v=3\sqrt{3}\lambda_{SO}$, as in the case of
the ZZ edge.
Unlike the SOC,
introducing the sub-lattice potential does not only change the bulk energy
spectrum, but also changes the edge-state dispersion. 
Nevertheless, the modification is quite small
and restricted only in the large $k$ region.
For small $k$, on the other hand, the edge-state dispersion is still intact
despite the parameter change.
As seen in \fig{fig:ac_dispersion}, the edge-state dispersion with a finite
sublattice potential shows only a very small deviation  
near $k=\pm\frac{\pi}{\sqrt{3}}$.

\section{The width of the edge state}
\label{sec:width}

\begin{figure*}[htb]
	\includegraphics[width=8cm]{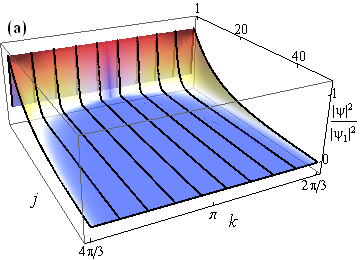}
	\includegraphics[width=8cm]{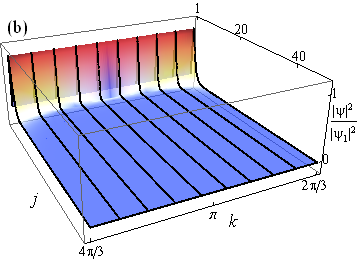} \\
	\includegraphics[width=8cm]{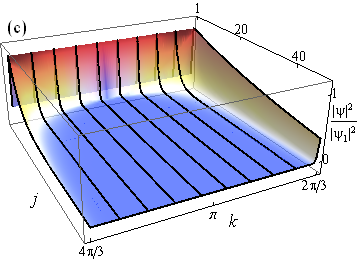}
	\includegraphics[width=8cm]{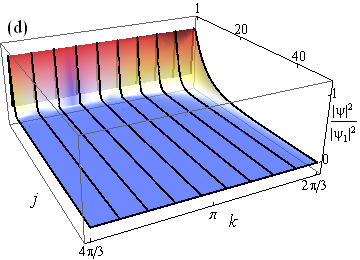}
	\caption{
	The wave function profile of the edge state of the ZZ edge as
	a function of the momentum $k$ in $x$-direction (along the edge)
	and the position $j$ in
	$y$-direction (perpendicular to the edge).
	Here, the values of
	SOC and the sublattice potential are (a) $\lambda_{SO}=0.05t$ and
	$\lambda_v=0$, (b) $\lambda_{SO}=0.15t$ and $\lambda_v=0$,
	(c) $\lambda_{SO}=0.05t$ and $\lambda_v=0.2t$, and
	(d) $\lambda_{SO}=0.15t$ and $\lambda_v=0.6t$.
	\label{fig:zz_wf}
	}
\end{figure*}

\begin{figure*}[htb]
	\includegraphics[width=8cm]{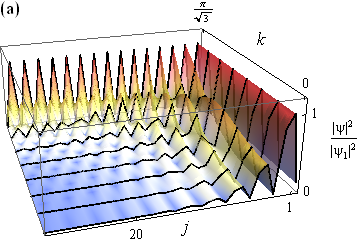}
	\includegraphics[width=8cm]{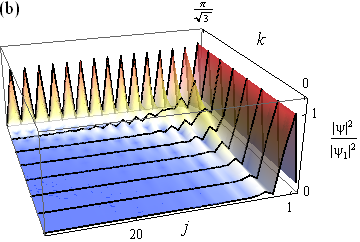} \\
	\includegraphics[width=8cm]{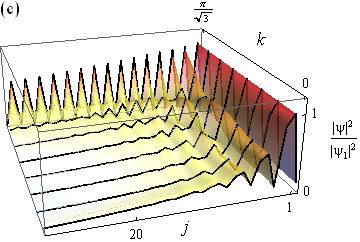}
	\includegraphics[width=8cm]{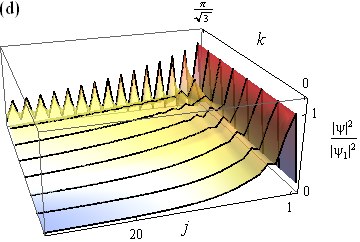}
	\caption{
	The wave function profile of the edge state with the AC edge as
	a function of the momentum $k$ in $x$-direction (along the edge)
	and the position $j$ in
	$y$-direction (perpendicular to the edge).
	Here, the values of the SOC
	and the sublattice potential are (a) $\lambda_{SO}=0.05t$ and
	$\lambda_v=0$, (b) $\lambda_{SO}=0.15t$ and $\lambda_v=0$,
	(c) $\lambda_{SO}=0.05t$ and $\lambda_v=0.2t$, and
	(d) $\lambda_{SO}=0.15t$ and $\lambda_v=0.6t$.
	\label{fig:ac_wf}
	}
\end{figure*}

In this section, we investigate spatial behaviors of the edge-state wave function.
The edge state appears only when we introduce an edge,
and its wave function is expected to be confined at a finite region 
near the edge.
The decaying factor, $\kappa$,  in Eq.~\eqn{eqn:decaying_factor}
shows that the wave function decays exponentially as it smears into
the inside of the bulk.
Figures \ref{fig:zz_wf} and \ref{fig:ac_wf} show the square of the
wave-function amplitude of the spin-up edge state in \eqn{eqn:general_sol}
as a function of the momentum
in $x$-direction along the edge
and the real space index $j$ in $y$-direction
perpendicular to the edge
for the ZZ edge and the AC edge, respectively.
In both cases, the edge-state wave functions are rather strongly localized
near $k=\pi$ for the ZZ edge
and $k=0$ for the AC edge.
These states gradually evolve to delocalized states as the momentum moves away
from the values mentioned above
until their energy dispersions submerge to the bulk energy
spectra at $k=\frac{2}{3}\pi$ and $\frac{4}{3}\pi$ for the ZZ edge
and $k=\pm\frac{\pi}{\sqrt{3}}$ for the AC edge.

For the localization properties of the edge-state wave function,
we can define a characteristic length scale 
from the decaying factor
\beq
\xi = 1/[\re~\kappa].
\eeq
This can be denoted as the spatial width of the edge state.
The decaying factors are determined by solving the coupled equations 
\eqn{eqn:zz_energy} and \eqn{eqn:zz_bc} for the ZZ edge,
and \eqn{eqn:ac_energy} and \eqn{eqn:ac_bc} for the AC edge.
Each edge-state wave function has 
two decaying factors for the ZZ edge and four decaying factors
for the AC edge.
Although the physical length scale is actually determined
by the smallest decaying factor
which gives the longest length scale,
we also consider the larger ones since 
they are useful in analyzing
the dependence of the width on the external parameters like
the bifurcation behavior as mentioned in our earlier work.\cite{Doh2013prb}

\fig{fig:zz_wf} shows the wave function of the edge state near the ZZ
edge as a function of the momentum $k$ in $x$-direction and the
position $j$ in $y$-direction.
The detailed analysis for the ZZ edge was studied in
our previous work,\cite{Doh2013prb}
where the edge-state width has two length scales on the ZZ edge.
The two widths are the same and almost constant near $k=\pi$ as the momentum varies. 
The widths do not change significantly until
the momentum difference from $\pi$ exceeds a certain value which
is a function of $\lambda_{SO}$.
Right after the momentum exceeds the value,
the two widths split into different values.
One of them decreases and the other monotonically increases and diverges when
the edge-state energy merges into the bulk energy.
The splitting of the widths forms a bifurcation behavior.

The spatial wave function profiles of the edge state with the AC edge are shown
in \fig{fig:ac_wf}.
The edge state is delocalized when the momentum reaches $k=\frac{\pi}{\sqrt{3}}$ where
the energy of the edge state merges into that of the bulk.
The largest one of the four edge-state widths monotonically increases
as the momentum increases up to 
$k=\frac{\pi}{\sqrt{3}}$ without any bifurcation behavior or
a constant-value region
unlike the case of the ZZ edge.

\begin{figure}[htb]
	\includegraphics[width=8cm]{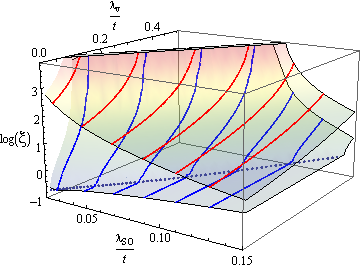}
	\caption{
	The widths of the edge states at $E=0$
	as a function of the SOC and the sub-lattice potential. 
	The upper surface with red lines is for the AC edge,
	and the lower one with blue lines for the ZZ edge.
	The blue dotted line denotes the bifurcation points
	of the widths of the ZZ edge state.
	\label{fig:edge_width}
	}
\end{figure}

The difference between the ZZ edge and the AC edge can be more predominent
when we change the strength of SOC.
The SOC dependence of the width for the ZZ edge 
at the momentum $k=\pi$ can be expressed in a simple
form\cite{Cano-Cortes2013prl,Doh2013prb}
\beq
\xi_\pi
=a\left[\mbox{arcsinh}\frac{t}{4\lambda_{SO}}\right]^{-1}.
\eeq
This expression is still valid with the sublattice potential, which only changes
the imaginary part of the complex decaying factor.\cite{Doh2013prb}
This shows the localized edge state on the ZZ edge actually gets delocalized
as the SOC strength increases.
On the other hand, increase of the
SOC strength enhances localization of the edge state on the
AC edge. It is clear when we compare Figures \ref{fig:ac_wf} (a) and (b), where
the wave function is more squeezed to the edge with the larger SOC strength.
According to the numerical results of a nano-ribbon honeycomb lattice,
the edge-state width of the AC edge is inversely
proportional to the bulk energy gap.\cite{Ezawa2013prb}
Since the gap is roughly proportional to the SOC strength for the small SOC,
the edge-state width of the AC edge decreases as the SOC increases.
This is summarized in \fig{fig:edge_width}, which shows that the edge-state width
of the AC edge
monotonically decreases as SOC, $\lambda_{SO}$, increases,
while that of the ZZ edge increases.

The broadening of the edge state on the ZZ edge due to the increase of SOC
seems counter-intuitive
in the sense that SOC develops
topologically nontrivial gap in the bulk.
Considering the band structure of the Kane-Mele model, however, SOC does not only 
intensify the topological nature of the system, but also modifies
the whole band structure.
Without the SOC and the sublattice potential,
the honeycomb lattice is semi-metallic with Dirac
cones
and shows a non-dispersive
localized edge state on the ZZ edge like
graphene.\cite{Fujita1996jpsj,Nakada1996prb}
In the strong limit of SOC, we can ignore the nearest-neighbor
hopping and the sublattice 
potential term.
This makes the system topologically trivial and asymptotically metallic.
Therefore, the increase of SOC in the 
Kane-Mele model does not always intensify the TI
state.
This explains why the metallic edge state finally delocalizes in the large limit of
SOC.

Unlike the SOC,
increasing the sublattice potential always drives the system to a topologically
trivial insulating state. The transition between the topologically non-trivial to
the trivial state should encounter a metallic state
due to the topological discontinuity.
The edge-state widths also represent this transition. 
As seen in \fig{fig:edge_width},
the edge-state widths on both of the two edges increase
as the sublattice potential increases,
and finally diverge when the system
becomes metallic at $\lambda_v=3\sqrt{3}\lambda_{SO}$.
Unlike the monotonic increase of the edge-state width of the AC edge, however,
that of the ZZ edge shows a transition behavior
from a SOC-insensitive
state to a SOC-sensitive state as the sublattice potential varies.
This originates from the bifurcation of the edge-state width on the ZZ edge.
The bifurcation points are denoted as a dotted line in \fig{fig:edge_width}.

\section{Edge-state gap for finite system}
\label{sec:finite}
Although our research is conducted in a semi-infinite lattice,
the results
is more important in a small system whose size  is comparable 
to the edge-state width.
\fig{fig:edge_width} shows that the edge-state width is smaller on the ZZ edge
than on the AC edge in nearly entire range of the parameters.
If we consider a finite width ribbon of the honeycomb lattice,
the metallic edge state of the two dimensional TI system with a finite width
is more favorable with the ZZ edge
where the edge-state width
is small and rather robust to perturbations such as the sublattice potential.

A ribbon with the AC edge can be useful when we try to control
the edge-state gap with the sublattice potential. 
For example, if the honeycomb lattice has a buckled structure
having the two sublattices on different planes
like silicene, the sublattice potential can be controlled by an external electric field.
Since the edge-state width of the AC edge is rather sensitive
to the sublattice potential, 
one can open a band gap at the edge state with a large external electric field.

\section{Summary}
\label{sec:summary}

We derive analytic equations for the edge state of the Kane-Mele model in
the semi-infinite honeycomb lattice with a ZZ edge and with an AC edge, respectively.
The analytic equations
are solved iteratively
for the energy and
the wave function of the edge state.
Our results have no size effects which is inevitable in the numerical calculation 
of a finite width ribbon.
From the analytic form of the wave function of the edge state, we define
an edge-state width which is a spatial decaying length of the edge-state
wave function perpendicular to the edge.
The calculated results of the edge-state width show peculiar behaviors and
dependencies on the SOC, the sublattice potential, and the
edge types.
The localized edge state on the ZZ edge is rather insensitive to the sublattice
potential. This gives robust nature of the metallic edge state
in a finite sized system.
On the other hand, the edge state of the AC edge is easily controllable with
the external parameters. Therefore, the edge-state gap
of the AC edge can be {\em turned on} with the sublattice potential in a finite
sized system.
Although the edge state of the TI is protected by the topology regardless
of the edge type,
our results show that the detailed dependence on the edge type can be crucial in
small-device manufacturing.

\begin{acknowledgments}
This work was supported by NRF of Korea (Grant No. 2011-0018306, HD and HJC)
and (Grant No. 2008-0061893, GSJ).

\end{acknowledgments}
\appendix
\section{Fermi velocity of the edge state on the ZZ edge
near $k=\pi$ without sublattice potential.}
\label{sec:zz_Fermi_vel}

In this appendix, we derive the Fermi velocity of the edge state on the ZZ edge
at $k=\pi$ by expanding
the coupled equations \eqn{eqn:zz_energy} and \eqn{eqn:zz_bc} near $k=\pi$.
If we ignore the sub-lattice potential $\lambda_v$ 
for the edge state on the ZZ edge,
the eigenvalue equation \eqn{eqn:zz_energy} can be simplified as 
\beq
\mathcal{E}^2=\left(\cos q-\cosh \kappa\right)^2
+t'^2\left(4\cos^2 q+4\cos q\cosh\kappa+1\right)
\eeq
where $q=k/2$,
\beqa
\mathcal{E} &\equiv & \frac{E}{4\lambda_{SO}\sin q},~ ~ ~
\mbox{and} \\
t' &\equiv & \frac{t}{4\lambda_{SO}\sin q}.
\eeqa
This is the second order equation of $\cosh \kappa$.
The two solutions for
$\cosh\kappa$
are 
\beqa
\nonumber
\cosh\kappa&=&\left(1-2t'^2\right)\cos q \\
&&\pm i\sqrt{t'^2-4t'^2\left(t'^2-2\right)\cos^2q
-\mathcal{E}^2},
\label{eqn:cosh_sol}
\eeqa
which are complex conjugates of each other.
Therefore, the boundary condition \eqn{eqn:zz_bc} can be written as
\beqn
\im\left[\left(3\cos q-\mathcal{E}\right)\cosh\kappa
+\left(\cos q-\mathcal{E}\right)\sinh\kappa\right] \\
=\im\left[\sinh\kappa\cosh\kappa^\ast\right]
\eeqn
where $\im[z]$ is the imaginary part of $z$.
Here $\kappa$ is generally a complex number, 
\beq
\kappa = \alpha+i\beta,
\eeq
where $\alpha$ and $\beta$ are real numbers.
From the solution for $\cosh\kappa$, \eqn{eqn:cosh_sol},
the real numbers, $\alpha$ and $\beta$ satisfy 
\beqa
\cosh\alpha\cos\beta &=& \left(1-2t'^2\right)\cos q, \\
\sinh\alpha\sin\beta &=& t'\sqrt{1-4\left(t'^2-2\right)\cos^2 q-
\left(\frac{\mathcal{E}}{t'}\right)^2}.
\eeqa
The boundary condition is rewritten in terms of $\alpha$ and $\beta$ like
\beq
\left(3\cos q-\mathcal{E}\right)\sinh\alpha 
+\left(\cos q-\mathcal{E}\right)\cosh\alpha 
=\cos\beta.
\label{eqn:analytic_bc_eqn}
\eeq

Now we expand $k$ near $\pi$,
by replacing $q$
by
$\frac{\pi}{2}+\delta q$.
If we expand $\delta q$ up to the first order, we get the followings
\beqa
\sinh\alpha &\simeq&  t', \\
\cosh\alpha &\simeq&  \sqrt{t'^2+1}, \mbox{ and} \\
\cos\beta   &\simeq&  \frac{2t'^2-1}{\sqrt{t'^2+1}}\delta q.
\eeqa
After inserting these into \eqn{eqn:analytic_bc_eqn},
we get 
\beqa
\nonumber
\mathcal{E}&=&\frac{(3\sinh\alpha+\cosh\alpha)\cos q-\cos\beta}
{\sinh\alpha+\cosh\alpha} \\
&\simeq&
-3\frac{t'}
{\sqrt{t'^2+1}}\delta q.
\eeqa
Since $\delta q = \frac{k-\pi}{2}$,
the energy dispersion of up-spin near $k=\pi$  on the edge is
\beq
E \simeq - v_F(k-\pi)
\eeq
where $v_F$ is the Fermi velocity at $k=\pi$
with the following form
\beq
v_F = \frac{6\lambda_{SO}t}{\sqrt{t^2+16\lambda_{SO}^2}}.
\eeq


\section{Energy dispersion of the edge state on AC edge
without sublattice potential.}
\label{sec:ac_energy_disp}

In this appendix, we will find 
the edge-state energy dispersion
on the AC edge
by showing that our trial solution for
the dispersion
satisfies Eq.\ \eqn{eqn:ac_energy} and \eqn{eqn:zz_bc}.
Our trial solution for 
the energy dispersion is 
$E_{k} = \pm t\sin \frac{\sqrt{3}}{2}k$
which is the envelope function of the bulk energy spectrum for $\lambda_{SO} = 0$.
If we use this function for the dispersion,
the eigenvalue equation \eqn{eqn:ac_energy} is reduced to
\beq
4\sigma\lambda_{SO}\sinh\kappa\left(\cosh\kappa-\cos q\right)
=\pm t\left(\cos q + 2 \cosh\kappa\right)
\label{eqn:ac_red_eig}
\eeq
where $q=\frac{\sqrt{3}}{2}k$, and $\sigma$ is $\pm 1$ depending on the spin.
If we  choose the positive sign in front of the nearest-neighbor hopping $t$
in the eigenvalue equation \eqn{eqn:ac_red_eig},
the edge-state wave function for the spin-up electron
has
negative $\kappa$ and is not confined near the edge.
One the other hand, if we choose
the negative sign for the dispersion of the spin-up electron as $E_{q}=-t\sin q$, 
$\kappa$ is positive and the eigenvector is 
\beqa
\Phi_\nu
&=&\bmat{c}
t\left(1+2 e^{iq}\cosh \kappa_\nu\right) \\
it\left(\cos q + 2\cosh\kappa_\nu\right) +t \sin q
\emat
\nonumber\\
&=&t\left(e^{-iq}+2\cosh\kappa_\nu\right)\bmat{c} e^{iq} \\ i \emat.
\eeqa
With this eigenvector, the boundary condition \eqn{eqn:ac_bc}
can be reduced to
\beq
\left|\begin{array}{cccc}
e^{iq} & e^{iq} & e^{iq} & e^{iq}  \\
i & i & i & i \\
e^{-\kappa_1} e^{iq} & e^{-\kappa_2} e^{iq} & 
e^{-\kappa_3} e^{iq} & e^{-\kappa_4} e^{iq} \\ 
ie^{-\kappa_1} & ie^{-\kappa_2} & ie^{-\kappa_3} & ie^{-\kappa_4}  
\end{array} \right|=0,
\eeq
which is satisfied trivially because the first and the second rows
are linearly dependent.
Therefore,
the energy dispersion of the edge state of the spin-up electron is,
in fact,
$E_q=-t\sin q$.
\bibliography{KM.bib}

\end{document}